\documentclass{IEEEtran}
\usepackage{newtxtext, newtxmath, amsmath, overpic, url}
\usepackage{todonotes}
\usepackage{cite}
\usepackage{algorithmic}
\usepackage{graphicx}
\usepackage{textcomp}
\def\BibTeX{{\rm B\kern-.05em{\sc i\kern-.025em b}\kern-.08em
    T\kern-.1667em\lower.7ex\hbox{E}\kern-.125emX}}

\begin{document}
 \author{\IEEEauthorblockN{Bernd Ulmann},
 \IEEEauthorblockA{anabrid GmbH /
  FOM University of Applied Sciences,
  Frankfurt/Main, Germany,
  Email: ulmann@anabrid.com\\}
 \and
 \IEEEauthorblockN{Shrish Roy},
 \IEEEauthorblockA{anabrid GmbH,
  Frankfurt/Main, Germany,
  Email: roy@anabrid.com}}
 \title{Building a simple oscillator based Ising machine for research and
  education}
 \maketitle
 \begin{abstract}
  Oscillator based \textsc{Ising} machines are non-von-Neumann machines ideally 
  suited for solving combinatorial problems otherwise intractable on classic 
  stored-program digital computers due to their run-time complexity. Possible 
  future applications are manifold ranging from quantum simulations to protein 
  folding and are of high academic and commercial interest as well. Described 
  in the following is a very simple such machine aimed at educational and 
  research applications.
 \end{abstract}
 \section{Introduction}
  \label{sec:introduction}
  Computer science and mathematics offer various frameworks for categorizing 
  the complexity of problems. One such classification is NP-hardness, which 
  characterizes problems for which the computation of solutions is 
  computationally intensive, i.\,e., they cannot be obtained in polynomial time 
  on a deterministic machine. Such problems manifest across a broad spectrum of 
  disciplines, including physics, biology, engineering, and even recreational 
  games. The \textsc{Ising} problem, categorized as an NP-hard problem, is 
  derived from the \textsc{Ising} model in physics, originally investigated by 
  \textsc{Ernst Ising} in 1925 (see \cite{ising}). This model is mathematically 
  formulated as follows (\cite{wang_2019}):
  \begin{equation}\label{eq:ising_model_equation}
    H=-\sum_{i,j}J_{ij}\sigma_i \sigma_j-\sum_j h\sigma_j,
  \end{equation}

  In order to comprehend this, the above equation can be envisioned as a 
  magnetic structure or material comprising diminutive particles that exhibit 
  spin $\sigma_i\in\{-1,+1\}$ indicating either a downward or an upward spin. 
  The coupling coefficient $J_{ij}\in \mathbb{R}$ represents a spin-spin 
  interaction term between adjacent particles and an external field $h$ (in 
  many cases, this is simplified by omitting the second summation term in the 
  literature). Aggregating these values yields a Hamiltonian $H$ that 
  characterizes the energy of the entire system. The \textsc{Ising} problem 
  then inquires as to the spin configuration of these particles (given a 
  magnetic structure where the interaction term is specified) that would result 
  in the minimum energy for the entire system. Thus, 
  \eqref{eq:ising_model_equation} becomes an optimization problem. 

  It is established that NP problems (decision problems solvable by a 
  non-deterministic machine within polynomial time) can be transformed into an 
  NP-hard problem utilizing polynomial resources. In pursuit of a theoretical 
  solution for all possible NP problems, the following concept emerged: Use a 
  physical system inspired by the \textsc{Ising} model consisting of particles 
  with varying upward and downward spins. When presented with an \textsc{Ising}
  problem, the system is configured on the basis of the Hamiltonian 
  corresponding to the problem. It then solves this optimization problem
  naturally evolving towards the lowest energy 
  state of the Hamiltonian, i.\,e., it converges towards the ground state.
  Consequently, its output values achieve sum minimization. 

  The methodologies for physically building such machines differ, adopting 
  various techniques to compute values. Nevertheless, mathematically, they all 
  strive towards the same objective. These machines are known as 
  \emph{\textsc{Ising} Machines} (see \cite{mohseni}, \cite{bian}).

  One particular implementation of such a machine involves the utilization of a 
  network of mutually coupled oscillators. In this context, the system's 
  convergence towards a ground state is facilitated by the synchronization of 
  these coupled oscillators (\cite{pikovsky}), rather than spins interacting 
  with one another (see also \cite{chou_2019} or \cite{wang_2019}).

  These oscillators may achieve synchronization either in phase or out of phase 
  (antiphase). The \textsc{Ising} machine based on oscillators, as delineated 
  in this paper, exhibits antiphase synchronization; that is, two mutually 
  coupled oscillators will synchronize such that one oscillator is out of phase 
  with the oscillator it is interacting with. Consequently, max-cut problems 
  are identified as the optimal choice for this type of 
  architecture.\footnote{Apart from max-cut problems there is a plethora of 
  application areas for machines of this type, including protein folding 
  (\cite{ortitz}), quantum simulation (\cite{briggs}), cancer research 
  (\cite{wang_2019}), neuromorphic computing (\cite{effenberger}) and many 
  more. A comprehensive list of \textsc{Ising} formulations for a wide variety 
  of NP problems can be found in \cite{lucas}.}

  The max-cut problem is to find a bipartition of the vertices of an
  edge-weighted graph $G(V,E)$ maximizing the weight of the edges crossing the 
  partition. Here, $V$ is the number of vertices and $E$ is the number of edges.
  The implementation of max-cut within the \emph{Ising} model is relatively 
  straightforward, given that positive (antiferromagnetic) couplers correspond 
  to edges whose removal decreases the objective function value. Consequently, 
  the mapping is direct and aligns with the adjacency matrix of the respective 
  graph.

  To ensure that the oscillators will eventually lock in only two possible
  phase relationships, $0$ and $\pi$ respectively, a \emph{second harmonic
  injection locking} (\emph{SHIL}) signal is fed to all oscillators in such a 
  machine. This SHIL signal has twice the frequency with respect to the 
  resonance frequency of the individual oscillators.

  Given that if the edge weight between two vertices of a graph $G$ is 
  $\mu_{ij}$ then based on \eqref{eq:ising_model_equation}: $J=-\mu_{ij}$. 
  Consequently, the resulting adjacency matrix delineates the specific 
  oscillators that are to be interconnected. The system subsequently evolves 
  to the desired solution of the max-cut problem.
  
  It should be noted that
  the mapping of alternative optimization problems presents significant 
  complexities. The conversion of \emph{quadratic unconstrained binary 
  optimization} (\emph{QUBO}) problems (see \cite{kochenberger}) to 
  \textsc{Ising} formulations, and vice versa, has been meticulously examined 
  by Dwave (see \cite{Dwavesystems}), providing valuable insights into the 
  mapping of max-cut problems onto the hardware described in this paper. 

  Nevertheless, the precise mapping of generalized QUBO problems onto a 
  specific hardware is yet to be fully elucidated. 
 \section{Implementation}
  The system described in the following consists of just eight oscillators 
  with an all-to-all interconnect scheme and is capable of solving max-cut 
  problems as described above. Due to its simplicity this system is ideally 
  suited for spin off replicas and allow students the chance of getting actual 
  hands-on-experience with such novel computing approaches.

  The first task towards building a little demonstration system is the 
  selection of a suitable oscillator type such as ring oscillators, LC 
  oscillators, phase shift oscillators, etc. The selection of an 
  oscillator must also take into account the question if the synchronization
  connection should be a dedicated input or if synchronization and output 
  signal are connected to the same port of the oscillator. The former variant
  allows for asymmetric coupling matrices while the latter would imply 
  symmetric coupling matrices, which would have the advantage of only requiring
  a triangular coupling matrix instead of a full quadratic matrix for an 
  all-to-all coupling, thus subtantially decreasing the overall complexity of
  the machine.

\begin{figure}
   \centering
   \includegraphics[width=.48\textwidth]{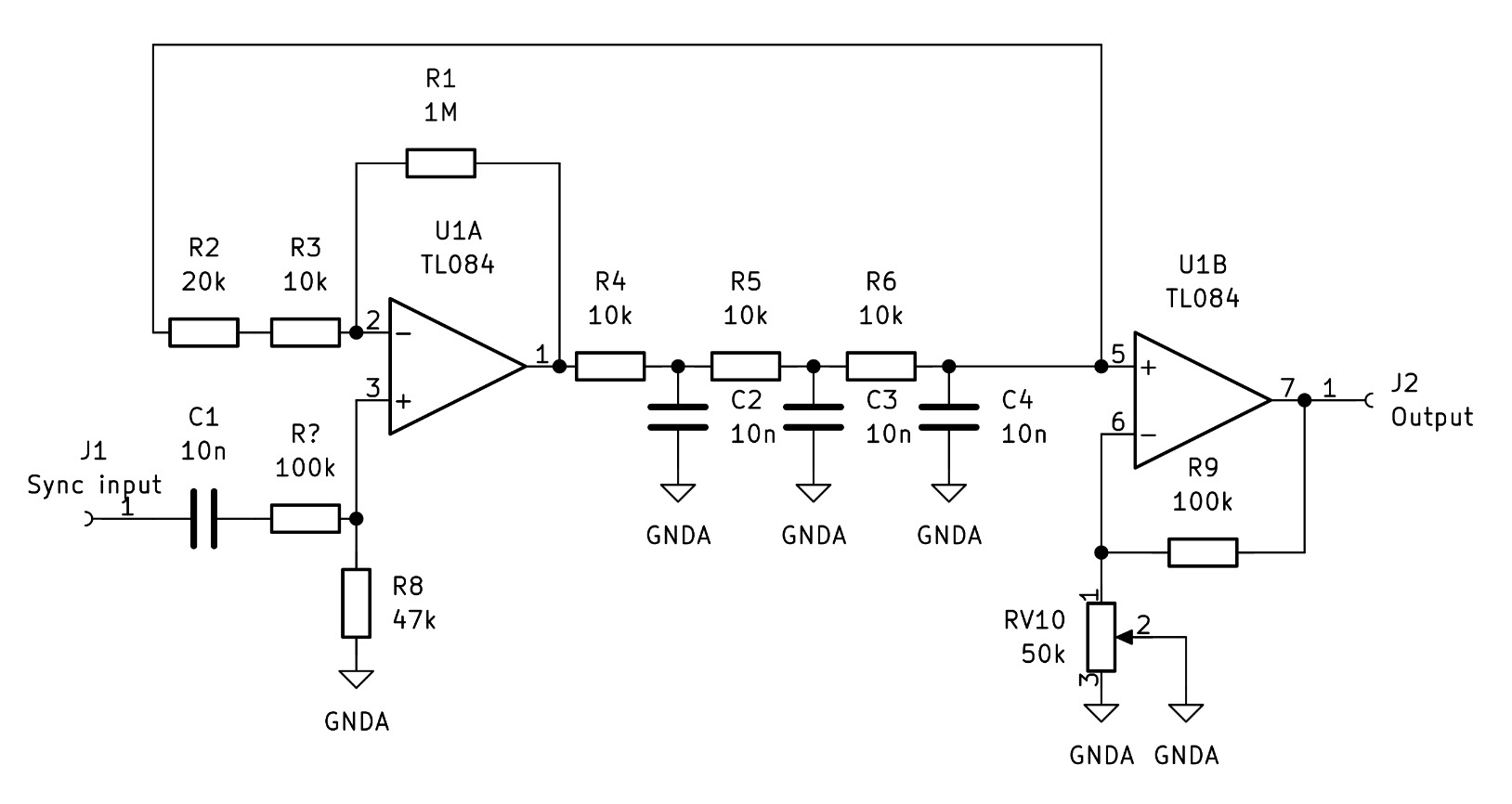}
   \vspace*{-4mm}
   \caption{Schematic of the phase shift oscillator}
   \label{pic_pso}
  \end{figure}

  To achieve utmost flexibility in the configuration of this experimental 
  machine it was decided to use oscillators with separate synchronization and 
  output ports, thus allowing asymmetric coupling schemes. Due to its inherent 
  simplicity and well-behavedness, a phase shift oscillator shown in figure 
  \ref{pic_pso} was chosen. The main operational amplifier on the left has the 
  synchronization input connected to its non-inverting input, thus assuring 
  that there will always be an inherent phase shift of $\pi$ between this input 
  and the output of the oscillator. The second operational amplifier on the 
  right just serves as an output buffer, the signal amplitude is set to 
  $\approx 4\text{V}_\text{pp}$ with \texttt{RV10} with the oscillator running
  from a symmetric $\pm15$V source.

  Eight such oscillators were built on two Euro format PCBs using cheap 
  TL084 quad operational amplifiers. The resonance frequency is about 3.8 kHz
  with the parts values shown in the schematic.

  To achieve all-to-all coupling, 56 coupling weights are required (there is 
  no need to couple an oscillator to itself). Each such weight consists of an
  AD5293 digital potentiometer with a resolution of $10$ bits. 24 such devices
  are contained on an anabrid Model-1 (\cite{ulmann_model1}) analog computer
  DPT24 module and can be configured by means of an attached digital computer 
  using a simple hybrid computer controller (\cite{ulmann_hc}), which also 
  contains eight such digital potentiometers. All in all two DPT24 and one HC
  module were needed to implement the required 56 digital potentiometers.

  It is mandatory that all connection weights representing a certain graph are
  activated simultaneously to avoid that the machine locks onto some spurious 
  solution. This is achieved by eight analog electronic switches, one dedicated
  to the synchronization input of each oscillator. A typical run of the machine
  follows these steps:
  \begin{enumerate}
   \item Deactivate all synchronization inputs.
   \item Set the coupling weights as required by the underlying graph.
   \item Let the oscillators run freely for a short moment to ensure random 
    phase relationships between the oscillators.
   \item Activate all synchronization inputs.
   \item Wait for a few oscillation periods (typically $<10$) until the 
    phase relationships have stabilized.
   \item Read out the phases representing the solution for the problem.
  \end{enumerate}

  These steps are performed by a simple control program as shown in figure
  \ref{pic_program}.
  \begin{figure}
   \centering
   {\scriptsize
    \begin{verbatim}
use strict;
use warnings;
use IO::HyCon;
...
my $ac = IO::HyCon->new($yml_filename);
$ac->setup();                   # Set all connection weights

my (%counter, $i);
for my $run (1 .. $runs) {
    $ac->digital_output(0, 0);  # Switch off sync inputs of the PSOs
    usleep($free_interval);     # Let the oscillators run freely
    $ac->digital_output(0, 1);  # Enable sync inputs
    usleep($locked_interval);   # Time for the oscillators and the 
                                # phase detectors to settle down 
    my $result = join('', @{$ac->read_digital()}[0..3]);
    print "($result) ";
    $counter{$result}++;        # Make result count
}
...
# Print statistics...
...
    \end{verbatim}
   }
   \vspace*{-5mm}
   \caption{Basic control program structure}
   \label{pic_program}
  \end{figure}

\begin{figure}
   \centering
   \includegraphics[width=.48\textwidth]{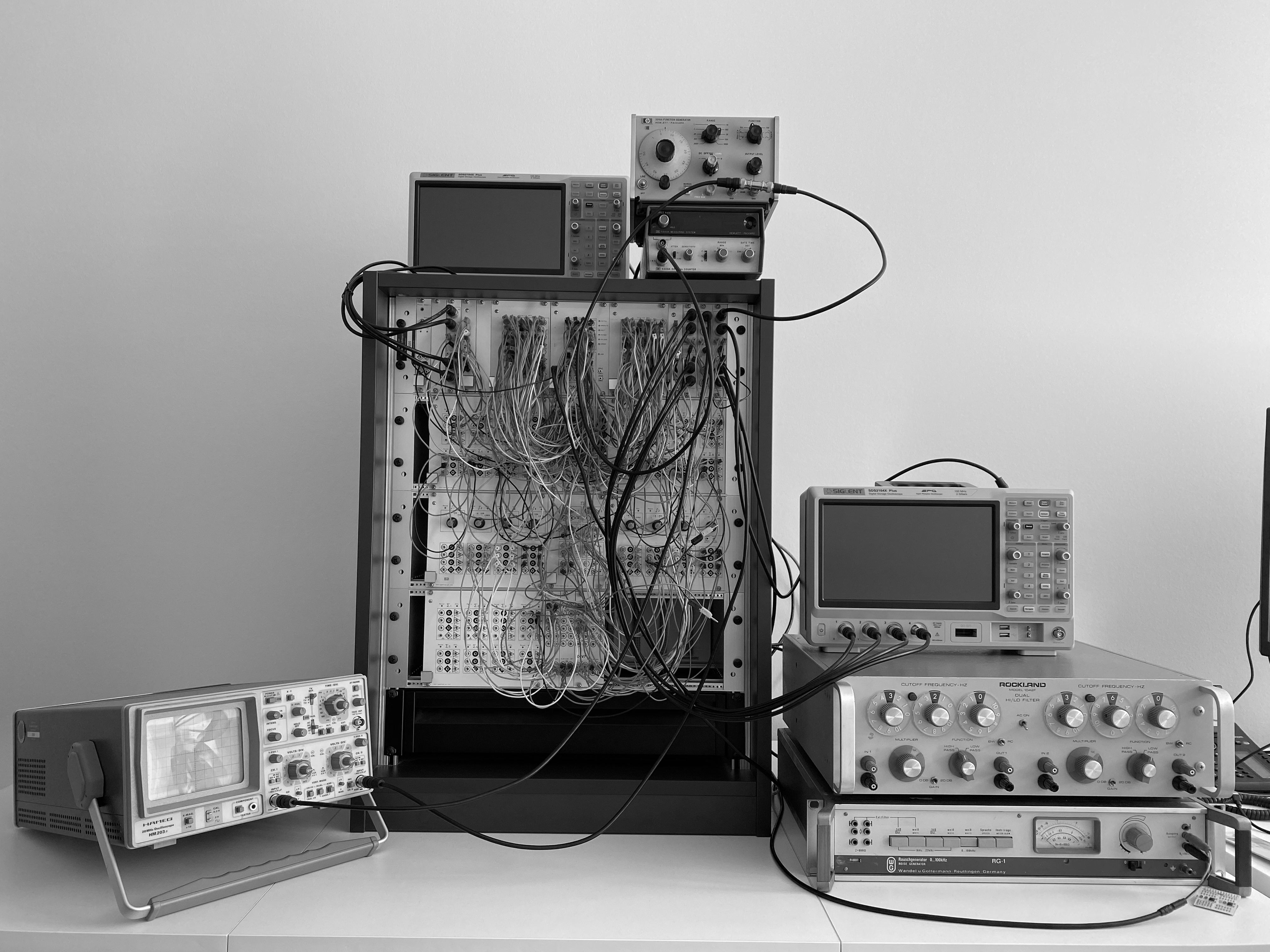}
   \caption{Setup of the eight oscillator \textsc{Ising} machine}
   \label{pic_8om}
  \end{figure}

  Figure \ref{pic_8om} shows the final setup of the system. The actual system
  is built from anabrid Model-1 analog computing modules which are connected 
  by patch cables. Results are displayed by means of two four channel digital
  oscilloscopes, one on top of the machine, one to its right. Also on top is 
  a HP 3310A function generator delivering the SHIL signal as well as a 
  frequency counter. The remaining equipment is used for various experiments
  exceeding the scope of this paper. 

\begin{figure}
   \centering
   \includegraphics[angle=90,width=.48\textwidth]{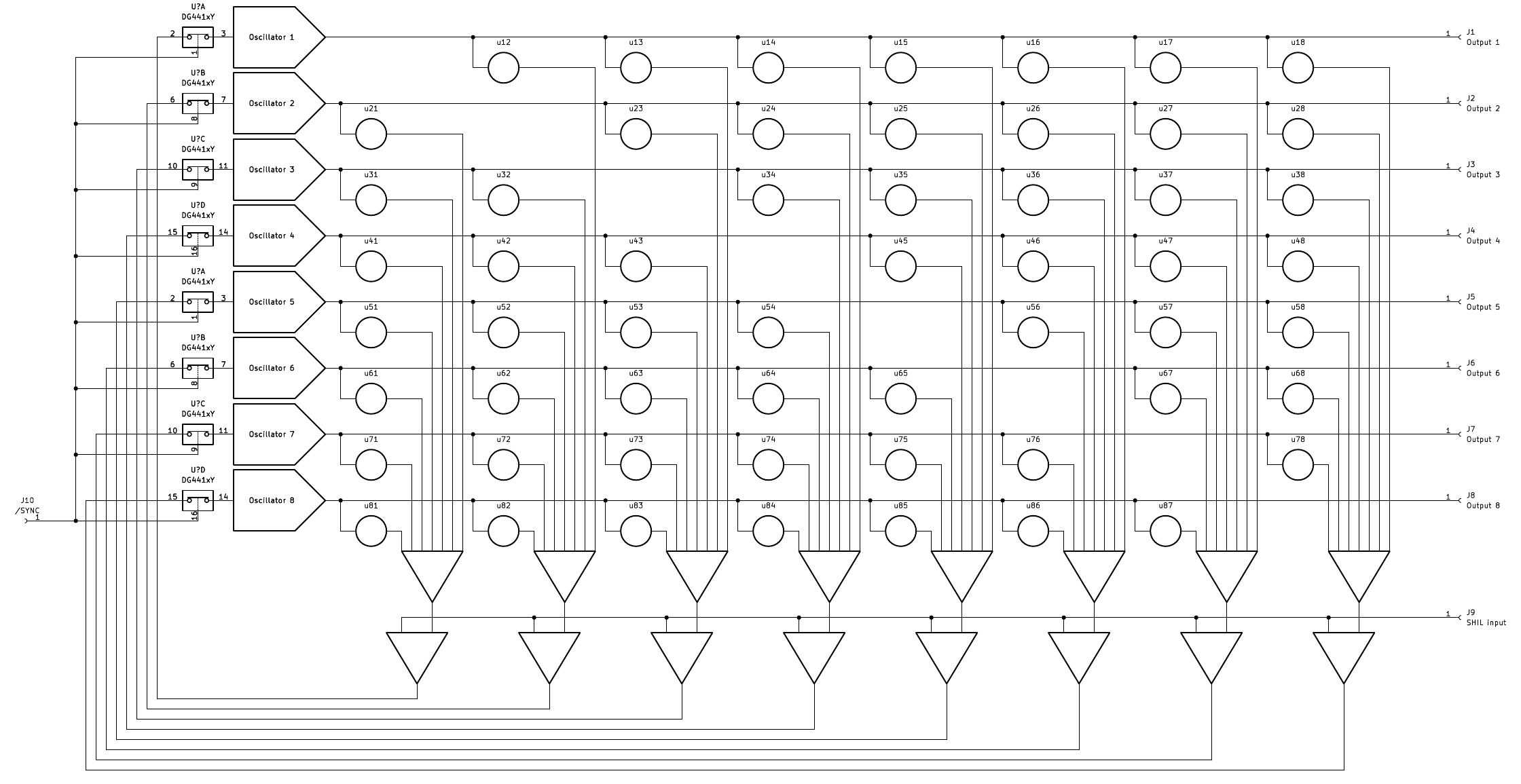}
   \caption{Schematic of the eight oscillator \textsc{Ising} machine}
   \label{pic_schematic}
  \end{figure}

  The schematic of the overall setup is shown in figure \ref{pic_schematic}. 
  The outputs of the eight oscillators are fed to a symmetric matrix consisting
  of $56$ coupling coefficients (the diagonal is $0$ as coupling an oscillator 
  to itself is not required) denoted by circles. The outputs of this matrix are
  summed columnwise by summers with seven inputs each (the triangular shapes 
  on the right). Since the summers used here are standard analog computer 
  summers, they perform an implicit sign reversal, so the output of each 
  column summer is fed to another summer whose main purpose is to revert this 
  sign inversion. These summers are also fed with the external SHIL signal. 
  The resulting eight outputs are then connected to the synchronization inputs
  of the oscillators by means of analog electronic switches controlled by an 
  external signal \texttt{/SYNC}. Using this signal all oscillators can be 
  switched from free running to coupled mode simultaneously.

  Not shown are the seven phase detectors used to read out the phase 
  relationships after the oscillators have settled down to a solution. These,
  too, were built from standard analog computer components. The basic idea is
  to multiply the output of an oscillator with the output of the reference 
  oscillator (arbitrarily chosen to be the first oscillator in the setup). 
  Since only two phase relationships of $0$ and $\pi$ (in practise deviations
  of several degrees from these ideal values are observed) are possible due to 
  the external SHIL signal, this product will be (roughly) 
  $\pm\sin(\omega t)^2$. Using an integrator with limiting diodes in its
  feeback path it can be distinguished between these two results. The outputs 
  of these phase detectors can then be read by the hybrid controller.
 \section{Results}
\begin{figure}
   \centering
   \includegraphics[width=.48\textwidth]{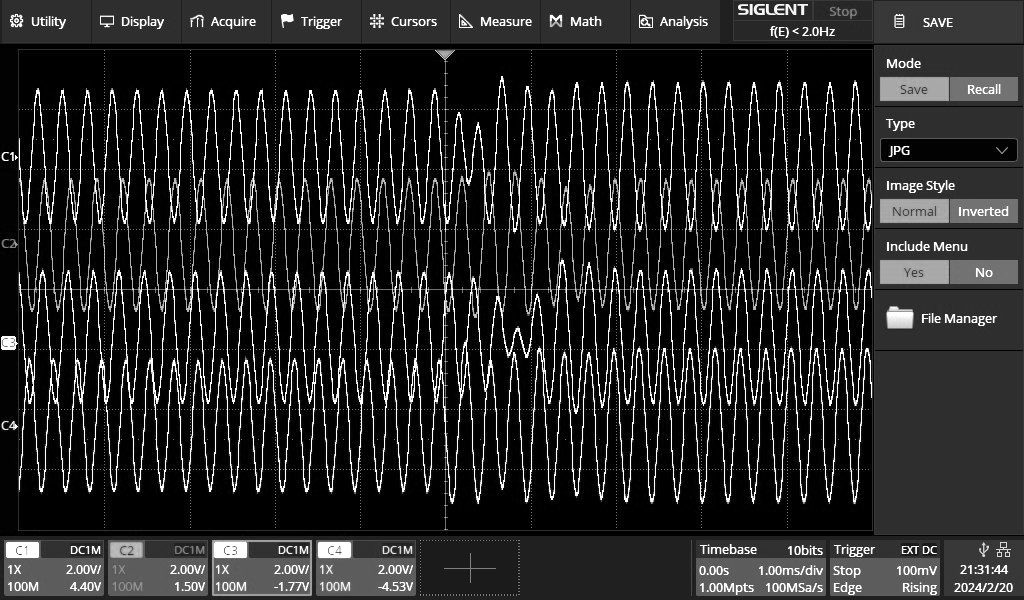}
   \caption{Typical behaviour of the system}
   \label{pic_result}
  \end{figure}

  The behaviour of the system can be seen in figure \ref{pic_result} (only four
  of the eight oscillators are shown). The little blue triangle in the middle of
  the screen denotes the trigger event. Before this, the oscillators are free
  running, while the coupling weights are applied instantaneously at the 
  trigger event. It can be seen that in this case a stable solution is obtained 
  after only seven periods of oscillation. 

  In setting up a problem on the machine, the non-zero weights of the coupling
  matrix were all set to the same value. It should be noted that the actual 
  non-zero value has a large influence on the performance of the system and
  the results obtained and varies from problem to problem depending on the 
  underlying graph. Values between $0.1$ and $0.3$ were sufficient for most 
  of the problems under consideration.

  Apart from a few exceptions, the machine yields correct solutions in mostly
  $100\%$ of runs. In cases where more than one correct solutions exists, the
  solutions obtained were typically not of equal or similar probability -- the 
  system seems to prefer some solutions over others. All in all the performance
  of this simple setup is pretty remarkable and will spur further research
  and experimentation.

  A number of topics has been identified which will be examined more thoroughly
  in the near future:
  \begin{itemize}
   \item How do different types of oscillators behave in an actual setup 
    (LC oscillators, ring oscillators, etc.)?
   \item How many bits of resolution are required for the coupling weights
    depending on various problem classes? For unweighted max-cut basically one 
    bit would suffice given that there is some global non-zero coupling weight 
    that can be controlled, but what about other tasks?
   \item Exploring different coupling topologies. As desirable as an all-to-all
    connectivity is it does not scale well with increasing numbers of 
    oscillators (at least $\approx 10^3$).
   \item An interesting question is that of the effects of noise on such a 
    system.
   \item If symmetric connection matrices are sufficient for all or at least
    the majority of problems, oscillators with combined synchronization input
    and output would allow to convert the connection matrix to a triangular 
    matrix, thus saving $50\%$ of the digital potentiometers.
  \end{itemize}
 \begin{IEEEbiography}[{\includegraphics[width=1in,height=1.25in,clip,keepaspectratio]{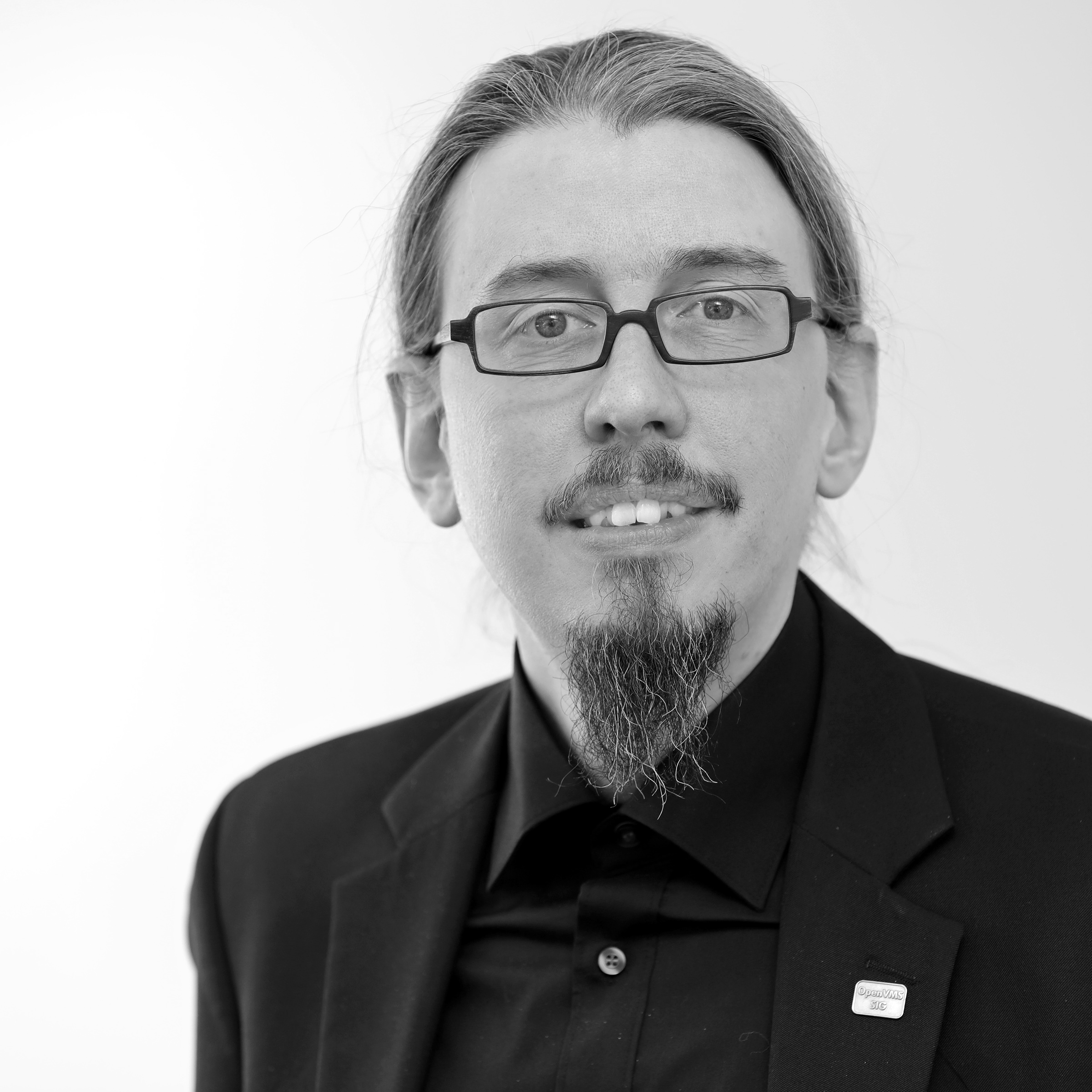}}]{Bernd Ulmann} was born in Neu-Ulm, Germany in 
  1970. He received his diploma in mathematics from the Johannes 
  Gutenberg-Universit\"at, Mainz, Germany, in 1996. He received his Ph.D. from
  the Universit\"at Hamburg, Germany, in 2009. 

  Since 2010 he is professor for business informatics at the FOM University of
  Applied Sciences, Frankfurt/Main, Germany. His main interests are 
  analog and hybrid computing, the simulation of dynamic systems and operator
  methods. He is author of several books on analog and hybrid computing.
 \end{IEEEbiography}
 \begin{IEEEbiography}[{\includegraphics[width=1in,height=1.25in,clip,keepaspectratio]{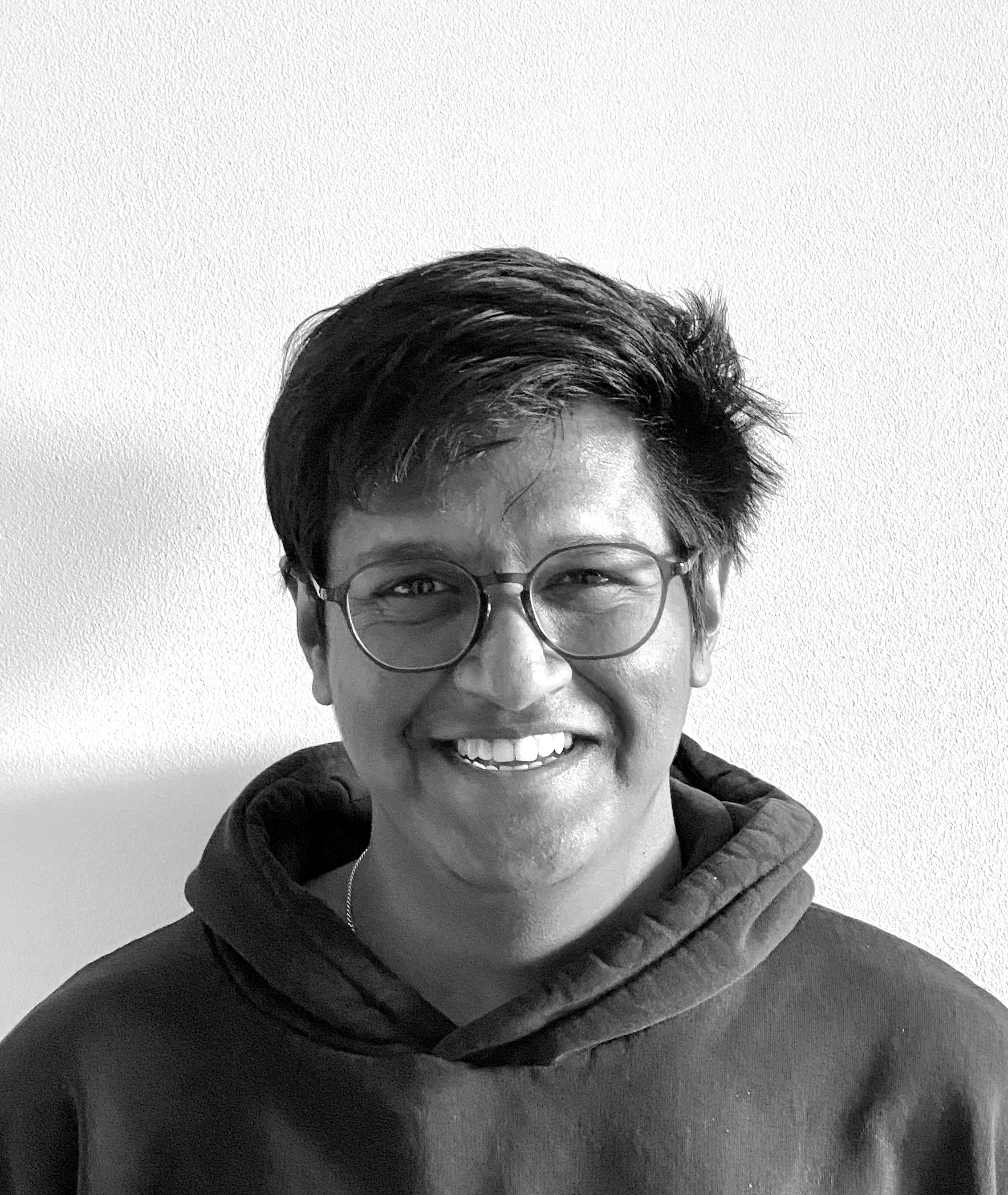}}]
  {Shrish Roy} was born in 2000 in Thane, Maharashtra, India. He completed his 
  master's degree in Mathematical Physics at the Eberhard Karls University of 
  T\"ubingen in 2024. Since January 2024, he has been working as a research 
  assistant at anabrid GmbH. His primary research interests include 
  unconventional computing methods, biologically inspired computing, and the 
  mathematics underlying information exchange in logical computation and its 
  physical manifestations.
 \end{IEEEbiography}
\end{document}